\documentclass[aps,prl,twocolumn,epsfig,groupedaddress,showpacs,floatfix]{revtex4}

\usepackage{epsfig}
\usepackage{amssymb,amsmath}
\begin{document}

\newcommand{\qs}{Q_{\rm sat}}
\newcommand{\qsa}{Q_{\rm sat, A}}

\def\Ntos{\smash{\mathop{\longrightarrow}\limits_{\raise3pt\hbox{$\sqrt{s}\to\infty$}}}}
\def\menor{\smash{\mathop{<}\limits_{\raise3pt\hbox{$\sim$}}}}
\def\maior{\smash{\mathop{>}\limits_{\raise3pt\hbox{$\sim$}}}}

\title{Percolation Effects in Very High Energy Cosmic Rays}
\author{J. Dias de Deus}
\email[Corresponding author]{, jdd@fisica.ist.utl.pt}
\author{M.C. Esp\'{\i}rito Santo}
\author{M. Pimenta}
\affiliation{CENTRA, LIP and IST, Av. Rovisco Pais, 1049-001 Lisboa, Portugal}
\author{C. Pajares}
\affiliation{Dep. Fisica Particulas, Univ. Santiago de Compostela, 
15706 Santiago de Compostela, Spain}

\date{\today}

\begin{abstract}
Most QCD models of high energy collisions predict that the inelasticity $K$ 
is an increasing function of the energy. 
We argue that, due to percolation of strings, this behaviour will
change and, at $\sqrt{s} \simeq 10^4$~GeV, the inelasticity will 
start to decrease with the energy.
This has straightforward consequences in high energy cosmic ray physics:
1) the relative depth of the shower maximum $\overline{X}$ grows faster
with energy above the knee; 
2) the energy measurements of ground array experiments at GZK energies
could be overestimated.
\end{abstract}


\maketitle


\noindent 
Most QCD-inspired models of multiparticle production predict, 
in hadron-hadron and nucleus-nucleus 
collisions at high energy, an increase with energy 
of the inelasticity parameter, 
$K \equiv 1 - x_F$,
where $x_F$ is the momentum fraction carried by the 
fastest particle: 
Multiple Scattering Model~\cite{multscat}, Dual and string Models~\cite{string}, 
and Minijet Model~\cite{minijet}. Recently, several papers appeared studying collective 
and non-linear QCD effects, based on the Colour Glass Condensate Model~\cite{colorglass}, 
Reggeon Calculus~\cite{reggeon}, and Strong Field string 
Model~\cite{strongfield1,strongfield2}. 
These models predict large stopping power and a decrease of the momentum 
fraction carried by fast particles.

Two aspects are essential in the String Percolation Model~\cite{percol1,percol2}
that we use here:
1) At low energy (or density) leading valence strings are produced. 
As the energy increases, energy is drawn from the valence strings to produce, 
centrally in rapidity, sea strings. As in all the models 
mentioned above, the inelasticity increases with the energy.
2) At very high energy, above the percolation threshold, percolation leads to the 
formation of a large cluster of strings and to the production of faster particles. 
As a consequence the inelasticity starts to decrease with the energy.

While the relatively low energy regime, with $K$ increasing, is similar to the models 
already mentioned, the higher energy regime, with decreasing $K$ and the regeneration 
of the fast particles, is new and has some straightforward consequences in cosmic 
ray physics. 


In the String Percolation Model~\cite{percol1,percol2} for hadron-hadron collisions, 
at low energy, valence strings are formed, forward and backward in the centre-of-mass, 
along the collision axis, containing most of the collision energy. 
As the energy increases, additional sea strings, central in rapidity, 
are created, 
taking away part of the energy carried  the valence strings. In the impact parameter 
plane all the 
strings look like disks, and we have to deal with a two dimension percolation problem.

The relevant parameter in percolation theory is the transverse density, 
$\eta$~\cite{percol3},
\begin{equation}
\eta \equiv {({r\over \overline R})^2} \overline N_s
\label{eq:eta}
\end{equation}
where $r$ is the transverse radius of the string, $\overline R$ 
the effective radius of the interaction 
area. $\overline N_s$, the average number of strings, depends on the density 
(centrality) and on the energy. The strings may overlap in the interaction area, 
forming clusters of $N$ strings. 
If $\eta \ll 1$, the average number of strings per cluster is $< N > \simeq 1$. 
If $\eta \gg 1, < N > \simeq \overline N_s$.

If $\overline n$ is the particle density for one string, $\overline m_T$ the average 
transverse mass produced from a single string, and there are $\overline N_s$ 
strings, one expects:
\begin{equation}
{dn \over dy} = F(\eta) \overline N_s \overline n \text {$~~~$and$~~$} 
< m_T > = {1 \over {\sqrt F(\eta)} } \overline m_T,
\label{eq:dndy&mt}
\end{equation}
with a colour summation reduction factor~\cite{percol4,percol5},
\begin{equation}
F(\eta) \equiv \sqrt{{1 -  e^{- \eta} \over \eta }},
\label{eq:F}
\end{equation}
The particle density does not increase as fast as $\overline N_s$ (this corresponds to 
the saturation 
phenomenon~\cite{percol2}), and $<m_T>$ slowly increases with energy and density. 
These features are seen in data (see, for instance, \cite{phenix}).

Following~\cite{percol6}, let us consider proton-proton collisions and write for the 
invariant s,
\begin{equation}
s \equiv (P_1 + P_2)^2 \simeq 4 P^2 = m^2 e^{\Delta Y}, 
\label{eq:s}
\end{equation}
where $\vec P_{1,2}$ are the momenta of the protons, 
$P = \mid\vec P_1\mid = \mid\vec P_2\mid$, $m$ is the proton mass, and $\Delta Y$ 
the length 
of the rapidity "plateau". For a string made up of two partons with Feynman-$x$ 
values $x_-$ and 
$x_+$, we have, for the centre-of-mass energy of the two partons, $s_1 = x_- x_+ s$. 
Assuming for simplicity a symmetrical situation around the 
centre-of-mass, $ x_- \simeq x_+ = \overline x$, the string centre-of-mass energy
is $s_1= \overline x^2 s$, and we can write the length of the rapidity plateau
for the string as: 
\begin{equation}
\Delta y_1 = \Delta Y + 2 ~ \ln ~ \overline x. 
\label{eq:dy1}
\end{equation}
If we write $ \overline x = 2 \overline p / \sqrt s$, where $\overline p$ is the 
absolute value of the momentum of the partons, we obtain:
\begin{equation}
\Delta y_1 = 2 ~  \ln ~ {2 \overline p \over m}.
\end{equation}
We shall assume, for sea strings, that $\overline p = {\rm constant}$, which implies
$\overline x \sim 1/ \sqrt s$, in agreement with the rise of parton density at small $x$.
Note that for valence strings, before energy degradation, $\overline p \sim P$,
and the full phase space is occupied by the valence strings.

If strings overlap in the interaction region, and if $<N>$ is the average number 
of strings per 
cluster we have, generalising~(\ref{eq:dy1}),
\begin{equation}
\Delta y_{<N>} = \Delta y_1 + 2 \ln <N>. 
\label{eq:dyn}
\end{equation}
At low energy/density $< N > \simeq 1$ and only short strings are formed, not 
contributing to cosmic ray cascades. At high energy/density 
$<N> \simeq \overline N_s$, percolation occurs and the situation 
changes.

In percolation theory the  average number $<N>$ of strings per cluster is related 
to the average 
area $<A>$, in units of $r^2$, occupied by a cluster~\cite{percol7},
\begin{equation}
<N> = <A> {\eta \over 1 - e^{- \eta}},
\label{eq:n}
\end{equation}
with $<A>$ given by~\cite{percol8}:
\begin{equation}
< A > = f (\eta)  \left[ ({\overline R \over r})^2 (1 -  e^{- \eta}) -1 \right]  + 1, 
\label{eq:A}
\end{equation}
where $f(\eta)$ is a percolation function, 
\begin{equation}
f (\eta) = \left(  1+  e^{-(\eta - \eta_c)/a}\right)^{-1}, 
\label{eq:f}
\end{equation}
and $\eta_c \simeq 1.15$ is the transition point and $a \simeq 0,85$ is a 
parameter controlling the slope 
of the curve at the transition point, with $f(\eta)$ changing from 0 to 1 at 
$\eta \simeq \eta_c$. 
We note that when $\eta \to 0, < A > \simeq 1$, and when $ \eta \to \infty,< A > 
\simeq (\overline R/r)^2$. 
This kind of parameterisation was tested in~\cite{percol7}.

The energy, in the centre-of-mass, carried by the produced particles from sea strings 
is given by:
\begin{equation}
E_{CM} =  
\int^{+ {\Delta y_{<N>} \over 2} }_{-{\Delta y_{< N >}\over 2}}  
< m_T > ~ {\rm coshy} ~ {dn \over dy} dy 
\end{equation}
and we obtain, making use of~(\ref{eq:dndy&mt})
and subtracting the 2 valence strings,
\begin{equation}
E_{CM} = \overline m_T \overline n {1 \over \sqrt{F(\eta)}} F (\eta)(\overline N_s -2) 
\left[e^{{\Delta y_{<N>} \over 2} }  
 - e^{-{ \Delta y_{<N>}\over 2} }  \right]. 
\label{eq:ecm}
\end{equation}
If we now require that asymptotically all the energy is carried by the percolating 
strings,
\begin{equation}
E_{CM} (\sqrt s \to \infty) = \sqrt s, 
\label{eq:ecmlim}
\end{equation}
we obtain, from~(\ref{eq:F}),~(\ref{eq:dyn}) and~(\ref{eq:ecm}),
\begin{equation}
\overline N_s \Ntos s^{\lambda}, \text{ with } \lambda = 2/7.
\label{eq:2/7}
\end{equation}
As $\overline N_s$ is proportional to the high energy bare Pomeron, the value of 
the intercept $\alpha_p$ is 
related to $\lambda$: $\alpha_p -1 = \lambda$.
This result is consistent with results from the Colour Glass Condensate 
Model~\cite{cgc}.

In order to have an estimate of the inelasticity $K$ we make use of the idea that, 
in the fragmentation 
of the string, produced particles are ordered in decreasing rapidity, and the fraction 
of momentum carried, 
relative to the momentum left, is always the same~\cite{krzy}.

At small $\sqrt{s}$, when the valence strings carry all the energy,
the fastest particle ($F$) is the leading particle ($L$) and 
$x_F = x_L = \alpha = {\rm const.}$, 
with $0 < \alpha \leq 1.$ When sea strings are produced, carrying an energy 
$E_{CM}$, we have,
\begin{equation}
x_L ={2 P_L \over \sqrt s} = \alpha \left( 1 - {E_{CM} \over \sqrt s} \right). 
\label{eq:xl}
\end{equation}
When the strings percolate, $E_{CM}\rightarrow 2 \mid\overline P\mid$ and for the 
fastest percolating particle ($P$) we have:
\begin{equation}
x_P = \alpha {< N > \over \overline {N_s}} {E_{CM} \over \sqrt s}. 
\label{eq:xp}
\end{equation}
As $E_{CM}$ is an increasing  function of $\sqrt s$, $x_L$ decreases 
with the energy and $x_P$ increases with energy. Thus,
\begin{equation}
K=\begin{cases}
1-x_L, & \text{ for } x_L > x_P \\
1-x_P, & \text{ for } x_L < x_P 
\end{cases}
\label{eq:k}
\end{equation}
%

In order to implement the model (equations~(\ref{eq:dndy&mt}), 
(\ref{eq:dyn}), (\ref{eq:ecm}) and (\ref{eq:xl}) to (\ref{eq:k})), 
we have to establish the relation 
between $\overline N_s$ and $\eta$(eq.~(\ref{eq:eta})), and to fix the parameters 
$r/ \overline{R}$,
$m_T$ and $\Delta y_1$. 
At some low energy threshold, 
${\sqrt s_t} \simeq 10$ GeV, we have just the valence strings and $\overline N_s=2$. 
At $\sqrt s \to \infty$, $\overline N_s \sim s^{\lambda}$
with $\lambda$ given by equation~(\ref{eq:2/7}). We then write:
\begin{equation}
\overline N_s = b + (2-b) ({s \over s_t})^{\lambda}.
\label{eq:ns}
\end{equation}
where the parameter $b=1.37$ was adjusted to agree with the data on $dn/dy$.
The remaining parameters were fixed to reasonable values: 
$r/ \overline{R}=0.2$, $m_T=0.78$ and $\Delta y_1=6$.
In this way, (\ref{eq:ecmlim}) was exactly satisfied.

\begin{figure}
\epsfxsize=6.cm
\centerline{\epsfbox{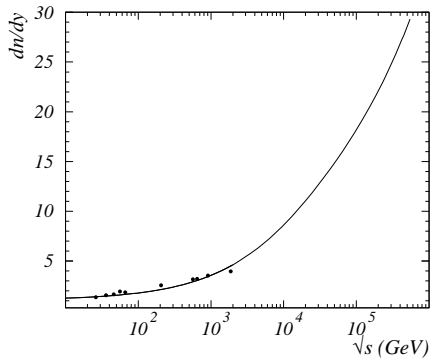}}
\vskip -0.3cm
\caption{
Particle density as a function of $\sqrt{s}$. Data points are from~\cite{dndy}.
The curve was obtained with~(\ref{eq:dndy&mt},\ref{eq:ns}) with $\sqrt{s_t}=10$~GeV and
$a=1.37$.
}
\label{fig:dndy}
\vskip -0.5cm
\end{figure}
In Fig.~\ref{fig:dndy} the $dn/dy$ data~\cite{dndy} is compared with the curve
obtained from~(\ref{eq:dndy&mt}) and~(\ref{eq:ns}). 
With the parameterisation for $\overline N_s$ (\ref{eq:ns}) we obtain that 
the critical density, $\eta_c$, occurs for $\sqrt s \simeq 10^4$ GeV.

\begin{figure}
\epsfxsize=6.cm
\centerline{\epsfbox{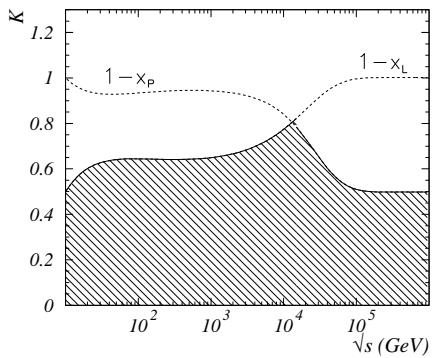}}
\vskip -0.3cm
\caption{
The inelasticity parameter $K$ as a function of $\sqrt{s}$, with $K=1-x_L$ at
relatively low energies and $K=1-x_P$ at energies above the percolation threshold.
}
\label{fig:k}
\vskip -0.5cm
\end{figure}
In Fig.~\ref{fig:k} we show the $\sqrt s$ dependence of the inelasticity $K$ 
(equation~(\ref{eq:k})). The behaviour of $1-x_L$ and $1-x_P$ is also shown in
the Figure. From the combination of the two curves, 
$K$ has a maximum at $ \sqrt s \simeq 10^4 $ GeV.


The development of showers, hadronic and electromagnetic, in cosmic ray physics, 
is critically 
dependent on the energy carried by the fast particles produced in the first hadronic 
collision, 
namely the leading particle. The shower evolves roughly exponentially, and 
reaches its maximum for a mean depth $\overline X$ (after the first collision). 

\begin{figure}
\epsfxsize=5.7cm
\centerline{\epsfbox{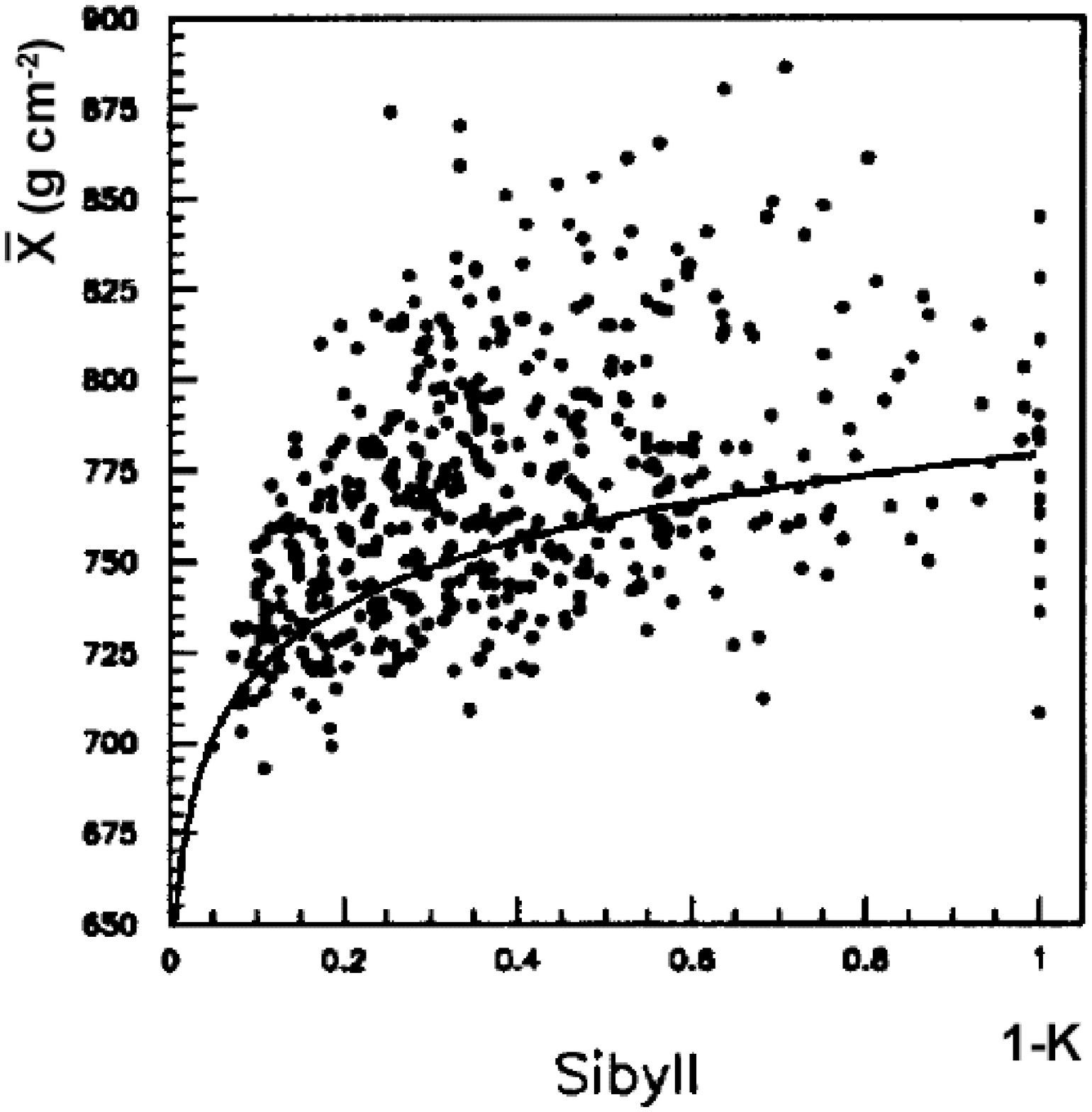}}
\epsfxsize=5.5cm
\centerline{\epsfbox{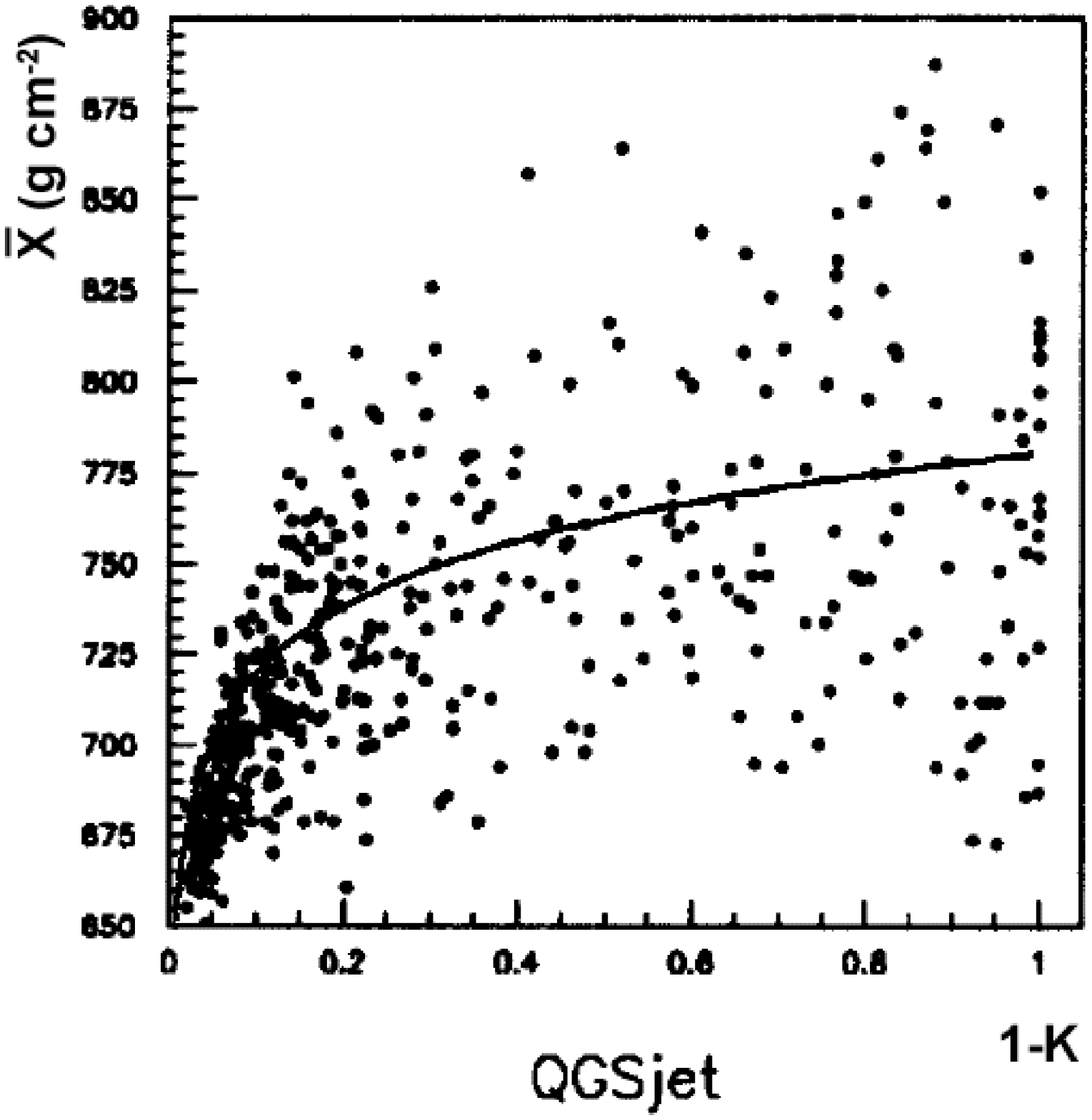}}
\vskip -0.3cm
\caption{
The relative depth of the shower maximum as a function
of the elasticity parameter $1-K$ is shown for Sibyll and QGSjet
simulations (taken from~\cite{pryke}) and compared to the
parameterisation of equation~(\ref{eq:X}).
}
\label{fig:scat}
\vskip -0.5cm
\end{figure}

Assuming now a simplified branching model, the relative shower maximum $\overline X$ 
can be expressed as:
\begin{equation}
\overline X = X_0 log_{10} \left[ ( 1 - K) E/E_0 \right], 
\label{eq:X}
\end{equation}
where $E$ is the laboratory energy ($E \simeq \frac{1}{2m}s$), 
$K$ is the inelasticity (as defined in 
the present percolation model). $X_0=60$~g/cm$^2$ and
$E_0=10^7$~eV are effective parameters related to the 
radiation length and to a low
energy threshold for the shower branching, respectively.
We have compared~(\ref{eq:X}) to simulations~\cite{pryke} using hadronic 
interaction generators (Sibyll~\cite{sibyll} and QGSjet~\cite{qgsjet}, 
based respectively on the Dual Parton Model and Quark Gluon string Model), incorporated
in the CORSIKA program~\cite{corsika}. 
The comparison is shown in Fig.~\ref{fig:scat}: the parameterisation~(\ref{eq:X})
is, at least for fixed energy, acceptable.

\begin{figure}
\epsfxsize=6.5cm
\centerline{\epsfbox{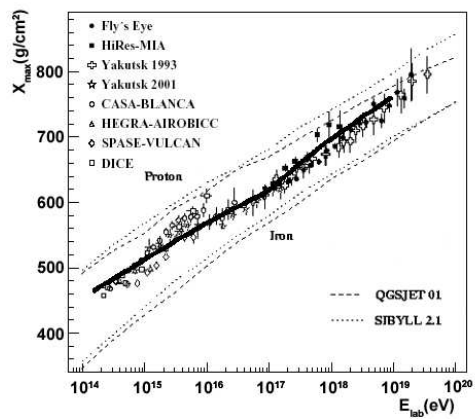}}
\vskip -0.3cm
\caption{
The relative depth of the shower maximum as a function of the primary energy.
The figure was adapted from~\cite{dova}, superimposing the 
result of the present percolation model (full line). 
Points are data and dashed lines show 
predictions of QGSJET and SIBYLL for protons and iron.
}
\label{fig:X}
\vskip -0.5cm
\end{figure}

The dependence of $X_{max} = \overline{X} + X_0$ on the primary energy $E$ is shown 
in Fig.~\ref{fig:X}. 
Near the percolation threshold ($E \sim 10^7$~GeV) there is a clear increase in
the slope of $\overline{X}(E)$. In the present percolation model,
the change in the slope of the $\overline{X}(E)$ curve can be qualitatively explained,
in a natural way, through a change in the behaviour of the inelasticity, due 
to the effect of percolating sea strings above a certain energy.
In contrast, this change in slope is usually explained by changing the fraction
of heavy nuclei in cosmic rays (see for instance ~\cite{dova}): 
this fraction would be higher
below the ``kink'' region (the region of fast growth of $K$), while above it 
the fraction of protons would rise (the region of decrease of $K$). In 
percolation models, a natural explanation arises without requiring a composition
change. In fact, Fig.~\ref{fig:X} shows that the percolation model line
naturally goes from the model predictions for iron to those for protons as the 
energy increases. A quantitative description of the data is however beyond 
the scope of the present work, as it would imply a dedicated Monte Carlo simulation
of proton-air interactions including percolation effects.

Furthermore, at very high energies a discrepancy between ground array experiments
and fluorescence detectors is usually quoted~\cite{watson}. We may note that the existence
of percolation will basically only affect the energy measurement of ground arrays,
which relies heavily on the Monte Carlo simulation of the shower development.
In fact, including percolation $K$ is lower and
fast secondaries are expected. $X_{max}$ will thus increase.
This will lead, at the atmospheric depth of AGASA (taking into account that the acceptance 
is maximal for relatively inclined showers) both to a larger total number of particles
and to a larger particle density in the region 600 m away from 
the shower core. This effect
could partially explain the apparent contradiction between ground
and fluorescence experiments at GZK energies. 

Additional work covering in detail the general case of AB collisions and the 
treatment of the high energy behaviour of cross-sections is in preparation \cite{paper2}.

We thank J. Alvarez-Mu\~niz and P. Assis for their enlightening comments.
We thank M.A. Braun, N. Armesto, F. del Moral, J.G. Milhano, C.A. Salgado and 
O. Catalano for discussions. 
J.D.D. and M.P. thank the Polytechnic Institute of Bragan\c{c}a and Prof.
Dion\'{\i}sio Gon\c{c}alves for the hospitality. J.D.D. also thanks the 
hospitality at Santiago.
This work has been partially done under contracts 
POCI/FIS/55759/2004 (Portugal),
FPA2002-01161 of CICyT of Spain FEDER funds from EU
and PGIDIT03PX1 C20612PN from Galicia.

\end{document}